\documentstyle[12pt]{article}
\newcommand{\doublespacing}{\let\CS=\@currsize\renewcommand{\baselinesstrech}
{2.0}\tiny\CS}

\begin{document}
\setlength{\baselineskip}{18.5pt}

\centerline{\LARGE{\bf Statistical Analysis of Quasar
    Data and Hubble Law }}

\vspace{0.9cm}

\centerline{\large{\bf  Sisir Roy$^{1,2}$ Dhrubajit Datta$^{3}$,
Malabika Roy$^{2}$ \& Menas Kafatos$^{2}$}}

\begin{center}
$^{1,3}${\bf Physics \& Applied Mathematics Unit 
\\Indian Statistical Institute \\Calcutta-700 035, India \\
e-mail : sisir@isical.ac.in \\
e-mail : {dhrubajit}-{datta}@yahoo.com }\\

\end{center}
\begin{center}
$^{2}${\bf Center for Earth Observing and Space Research \\
and  School of Computational Sciences\\ George Mason University\\
Fairfax, VA 22030-4444\\ USA.\\
e-mail : mroy@scs.gmu.edu \\
e-mail : mkafatos@gmu.edu}
\end{center}
\vspace{1cm}

\begin{center}
{\bf Abstract}
\end{center}
\noindent
\vskip5pt
The linearity of the Hubble relationship(i.e. between m and z) has been tested for galaxies and supernovea for low redshifts. We have studied this relationship for
quasars for data taken from Veron Cetti Catalogue(2003).The data from Veron Cetti Catalogue for quasars appear to be truncated. The data have been analyzed using various statistical methods which are suitable for analysing the truncated data. This analysis shows 
lineraity (in $\log z$)  of Hubble law for very small $z$ but non-linearity for high redshift.This will shed
new light not only on the quasar astronomy but also in the
cosmological debate.
\newpage

\section{\bf Introduction}
The general relationship between a distance of cosmic sources and corresponding  redshift allows
 one to establish some important properties
 for the universe which can be used to probe the spatial geometry and
 especially the underlying cosmological principles.  One of such relationships is  the m-z relationship between
 the apparent magnitude of the source and its 
redshift.
Astronomers normally work with quantities related to apparent
 magnitude $m$ and absolute magnitude $M$. The 
distance modulus is defined as the difference between $m$ and $M$ which is
 related to redshift $z$ and $H_0$, where $H_0$ 
 is the Hubble constant.
By analyzing the observational data, Hubble formulated a law which
 states that the galaxies appear to be receding with 
a velocity $v$ proportional to their distance $d$ from the observer:
$$v = H_0 d$$. This is known as the Hubble law, where $H_0$ is called the Hubble
constant. However, this relation can be 
derived from cosmological theory, if the universe is  assumed to be homogeneous
and isotropic. Various authors$^{1}$ 
discussed the limits of validity of the Hubble relation.
In the above Hubble relation, the distance $d$ should be very large so
that the recession velocity is 
larger than the radial component of the peculiar velocities : for
example this can be up to $1000 km s^{-1}$ 
for galaxies inside clusters and this puts a restriction , $d \geq
10h^{-1} $Mpc. This means the redshift $z $ has to be much greater than $10^{-2}$. 
However, the distance should not be so large that the recession velocity
 exceeds the speed of light. 
Crudely speaking, 
one can use the above relation for $d<< 300h^{-1} Mpc$,  or $z<<10^{-1}$. 
The distance $d$ can be shwon as $d \simeq \frac{c z}{H_0}\simeq 3000 h^{-1}$
Mpc for $10^{-2} \geq z \leq 10^{-1}$. 
This equation may be considered as first order approximation to the
formula for luminosity distance as a function of 
redshift $z$ in  the Friedmann model.
There are two luminosity functions i.e. absolute luminosity and
apparent luminosity. Normally astronomers work with 
absolute magnitude $M$ and apparent magnitude $m$, where  $d \simeq {m-M}$ is known
as distance modulus and is related to the 
luminosity distance.
Therefore, the relation between the distance modulus and $log z$ is considered to test Hubble law.
The linearity of Hubble relationship has been tested for Galaxies and
Supernova for low redshifts$^{2}$. Statistical 
analysis has been done on various samples for galaxies. 
Sometimes, the samples are constructed subjectively and
often samples are taken from Abell's$^{3}$ catalogue which assumes
Hubble's law as the selection criterion. Hoessel et al $^{2}$ took
samples of 116 galaxies from the Abell catalogue which supports Hubble
law. Segal and Nicoll$^{4}$ took infrared astronomical satellite
galaxy samples$^{5}$ and predicted an alternate redshift-distance
law i.e. $z \sim r^p$ where $p = 1,2,3$. Some other attempts$^{6}$
have been made with the IRAS $1.2$Jy Redshift survey. Efron and Petrosian$^{7}$
studied the viability of various statitical tests for truncated data in connection with redshift survey of galaxies and quasars. From the plot of redshifts $z_i$ and log luminosities (=$y_i$) for $210$ quasars they found the data as doubly truncated data.
Here, the trunctaion implies that it is not possible to get the information regarding the existence of ${(y_i,z_i)}$ if it fell outside of the region $R_i$ where, due to experimental constraints the distribution of each $y_i$ is truncated to a known interval $R_i$ depending on $z_i$. Truncated data may arise in various experiments. McLaren et al$^{8}$ analysed some experimental results in connection with Red Blood Cell volume distribution which lead to truncated data.
In section II we will describe nonparametric methods to find whether or not the 
apparent magnitudes $m$ are independent of redshifts $z_i$ for the trucated data taken from Veron Cetti catalogue for quasars. 
The cosmological implications are discussed in section III. 

\noindent 
\vskip5pt
\section{\bf Statistical Analysis of Quasar Data from Veron Cetti Catalogue }
\noindent
\vskip5pt
One of the main issues in analyzing astronomical data is to answer the following statistical question. Is a sample of observed points $(z_i, m_i)$ in the trncated data set of quasar survey consistent with the hyopthesis ${\bf H_0}$ that $z_i$ and $m_i$ are statistically independent ? Efron and Petrosian$^{7}$ investigated this issue in details using a small sub-sample of quasar data. Veron Cetti Catalogue$^{9}$ like other redshift survey provides a pair of measurements $(z_i,m_i)$. Various type of observational biases are ignored. One of the most common biases is  introduced by limiting $m_i$ of the survey.
We can write the data set as $(z_i,m_i)$ for $i = 1,2,3.....n$ with $m_i\leq m_0$.
The absolute magnitude $M_i$ can be estimated if we assume particular cosmological model.
The data set $(z_i,M_i$ can be reexpressed satisfying atruncated relationship
$$M_i \leq m_0 - 5 \log d + C$$ where $C$ is a constant which can be set to zero.
Efron et al anlyzed  the data from redshift survey of  492 galaxies and the magnitude limit $m_0 = 21.5$ of this survey leads to the truncated boundary 
$$M_i \leq 21.5 - 5 \log z_i$$.

The scatter plot of $$m \\{\rm vs}\\ \log z$$ 
hint that there is truncation in the data. Here, the idea of truncation is used in the sense that the observatiopns $$(z_i,m_i)$$ are observable if some condition or mathematical relation is satisfied say, $$\log z \geq a m + b$$, for some $a \& b$ .  It appears from scatter plot from Veron Cetti data that there exists at least one side turncation.  By considering $a = 3/7$ and $b= - 64/7$ we found that there are only $18$ data points among the $48683$ data points for which $\log z \geq 3/7 m - 64/7$. So we discard these $18$ data points and this number is negligible compared to the size of the data set and take $\log z \leq 3/7 m - 64/7$ as the truncation relationship. 
In the next step we will use the Test of Independence for truncated data as elaborated by Efron and Petrosian. 
Suppose the data consists of a random sample of $n$ pair from the joint distributions of
$${\rm data} = {(x_i,y_i), i = 1,2,....n}$$
For truncated data we assume that pairs $(x,y)$ are observable only if
they satisfy the truncation relationship 
$$y\leq u(x)$$ where $u(x)$ is a monotonic function of $x$.
Following Efron and Petrosian we took $x = m , y = -\log z , u(x) =
(-3/7)x + 64/7$. The test is to accept independence if $$|t_w{(\rm
data)}| \leq 1.96$$ and
 we take the rejection probability of the
permutation test to be approximately $0.05$. 
Here,  if we take $w_i = 1  \forall$ the test statistic 
 $$|t_w(\rm data)|= 704.162$$ with $$w_i =
\frac{x_i-x_{\rm min}}{u_i - u_{\rm min}}$$, which leads  locally the
most powerful test $$|t_w(\rm data)|= 875.594$$ . In both the cases 
the extremely large value of $t_w({\rm data})$
clearly rejects the hypothesis of independence.
 Here, $$ p-{\rm value}\\ \sim 0$$ where
 $$ p-{\rm value}$$
is the maximum level of significance under which null hypothesis
(here, independence of two variables) is accepted.
So $z$ and $m$ are not
independednt.

In the next step, we will try to get best fitting of these data using regression analysis.
Then we will investigate the conditons under which we can get Hubble relation.
The scatter plot of $z$ vs. apparent
magnitude $m$ is illustrated in Figure 1(a,b).
Our regression analysis shows that  we can use the following relation between m and z.
$$\log (m-12) = -4.528 + 16.542 {z}^{1/4} - 13.891 {z}^{1/2} + 3.884 {z}^{3/4}$$ for
 $ z \epsilon (0,7)$ i.e. for the whole range of $z$.

The following observation motivated us to analyse in a different manner.
Here, we observed that in the region ${[0.2950; 2.995]}$ , the truncation distribution of the  variable 
$$\frac{(m -12)}{(f(z) -12)}$$ can be well  approximated by beta distribution with parameters $a$ and $ß$ which are some functions of $z$. Precisely  given the truncation, the conditional distribution of  $$\frac{(m -12)}{(f(z) -12)}$$ can be well approximated by  beta$(a(z) , ß(z))$ where 
$$f(z) = (7/3)\log(z) + 64/3$$.Actually we did the above  analysis for $$z = 0.2950, 0.3050, 0.3150,…,2.9950$$. This information has been used to calculate the expected value of $m$ given $z$ for the last said values of  $z$. Then we went on doing usual regression analysis to find out $E (m|z) =19.484 + 0.886ln (z) - 0.783{(\log(z))}^ 2$ for the region ${[0.2950; 2.9950]}$. We found the $95\%$ tolerance interval with coverage probability $0.95$ in the similar fashion. $A \%$ tolerance interval with coverage probability $\alpha$ means that $A\%$ of the future observation will fall in the said interval with probability $\alpha$.In the specified region they are actually $(ml(z) ,mu(z))$ for given $z$ where 

$$m_u = 16.8 + 7.6263 z - 4.162 z^2 + 0.80 z^3$$
and
$$m_l = 12.51 + 5.576 z - 1.686 z^2$$.
They are shown in Fig.2. 
For the region $( 0; 0.2950)$ we use our general regression techniques to find out our prediction equation as  $m = 20.060 + 2.139\log z$ and prediction interval as

$$20.060 + 2.139 \log z ± 1.9631 \sqrt{0.4573}(1.0122 - 0.1132 \log z + 0.2937 {(\log z)}^2$$ 
Here, the prediction interval means that given z the value of m will fall in that interval with probability $0.95$.
\noindent
\vskip10pt
\section{\bf Possible Implications :}
It is found from our analysis for the data from Veron Cetti Catalogue that the Hubble relation between m and $\log z$ is valid for small $z$ i.e. for the range $$ z = [0 ; 0.295]$$. For higher values of $z$, we get different relation as found from regression analysis. Conventionally, the Hubble relation is explained as due to the Doopler mechanism for shift of the spectral lines. Now the deviation from Hubble relation may be due to some other mechanism for redshift. It may be pointed out that the environmental effetc for the quasars may be taken into consideration to explain thism deviation. This kind of environmental effect has been modeled in Doopler like mechanism considered in Dynamic Multiple Scattering (DMS)theory$^{10}$. This DMS is essntially based on the odea of correlation indeuced mechanism as discovered by Wolf$^{11}$. 
Finally we have plotted another curve in Fig.3 as $$V_{\rm effect}= V^* = m -M \\ {\rm vs}\\ z$$. This figure clearly indicates the existence of three different clusters of quasars. 
It is possible to identify these classes of quasars. The detail study of these clusters and its implications will be studied in a subsequent paper. 
 
\noindent
\vskip10pt
\section {References}
\noindent
\vskip5pt
\begin{enumerate}

\item Coles Peter and Lucchin Francesco (2002){\it Cosmology : The
    Origin and Evolution of Cosmic Structure}, 
2nd Edition, John wiley \& Sons, Ltd.p.77.
\item Abell G.O.(1958) Astrophys. J.Suppl. {\bf 3}, 211-288.

\item Hoessel J.G.,Gunn J.E. and Thuan T.X.(1980) Astrophys.J.{\bf 241},486-492.

\item Segal I.E.and Nicoll J.F.(1992)Procd.Nat.Acad.Sci.(USA){\bf 89},11669-11672.

\item Saunders W., Rowan-Robinson M. et al(1990)Mom.Not.R.Astron.Soc.{\bf 242},318-337.

\item Korany D.M and Strauus M.A.(1996) Tersting the Hubble Law with the IRAS 1.2 Jy redshift
      Survey, http://xxx.lanl.gov/astro-ph/9610034.

\item Efron B. and Pterosian V.(1992) Astrophysics Journal , {\bf 399}, 345-352.

\item McLaren C, Wagstaff M., Brittegram G., Jacobs A.(1991), Biometrics {\bf 47}, 607-708.

\item 10th Edition, Veron-Cetty M.P., Veron P., (2001)

\item Datta S., Roy S., Roy M. and Moles M., (1998), Phys.Rev.A,{\bf 58},720;\\
Roy S., Kafatos M. and Datta S.,(1999), Phys.Rev.A,{\bf 60}, 273.

\item Wolf E. and James D.F.V., (1996),Rep.Prog.Phys.,{\bf 59},771 
\end{enumerate}
\newpage
\noindent
\vskip5pt

\newpage
{\bf Figure Caption}

Figure 1(a). Scatter plot of z vs. m \\

Figure 1(b). Scatter plot of m vs. log(z) \\

Figure 2. Regressions for m vs z.\\

Figure 3. Scatter Plot of Veff vs z.

\end{document}